\documentclass{ws-ijbc}
\usepackage{epsfig}
\usepackage{ws-rotating} 
\usepackage{graphicx}
\usepackage{amsmath}
\usepackage{amsfonts}
\usepackage{amssymb}
\usepackage{subfigure}
\usepackage{psfrag}
\usepackage[english]{babel}
\usepackage[english]{varioref}

\begin{document}

\catchline{}{}{}{}{} 

\markboth{Matteo Sala, Cesar Manchein \& Roberto Artuso}{Estimating hyperbolicity of chaotic bidimensional maps}

\title{Estimating hyperbolicity of chaotic bidimensional maps}

\author{Matteo Sala$^*$}
\address{Center for Nonlinear and Complex Systems, Universit\`a dell'Insubria, Via Valleggio 11\\
Como, 22100, Italy\\
matteo.sala@uninsubria.it
}

\author{Cesar Manchein}
\address{Departamento de F\'isica, Universidade Federal do Paran\`a, 81531-980, Curitiba, PR, Brazil\\
Departamento de F\'isica, Universidade do Estado de Santa Catarina, 89219-710 Joinville, SC, Brazil\\
cmanchein@gmail.com
}

\author{Roberto Artuso\footnote{Istituto Nazionale di Fisica Nucleare, Sezione di Milano, Via Celoria 16, 20133 Milano, Italy}}
\address{Center for Nonlinear and Complex Systems, Universit\`a dell'Insubria, Via Valleggio 11\\
Como, 22100, Italy\\
roberto.artuso@uninsubria.it
}

\maketitle

\begin{history}
\received{(to be inserted by publisher)}
\end{history}

\begin{abstract}
We apply to bidimensional chaotic maps the numerical method proposed by Ginelli {\it et al.} to approximate the associated Oseledets splitting, i.e. the set of linear subspaces spanned by the so called {\it covariant Lyapunov vectors} (CLV) and corresponding to the Lyapunov spectrum.
These subspaces are the analog of linearized invariant manifolds for non-periodic points, so the angles between them can be used to quantify the degree of hyperbolicity of generic orbits; however, being such splitting non invariant under smooth transformations of phase space, it is interesting to investigate the properties of transversality when coordinates change, e.g. to study it in distinct dynamical systems. To illustrate this issue on the Chirikov-Taylor standard map we compare the probability densities of transversality for two different coordinate systems; these are connected by a linear transformation that deforms splitting angles through phase space, changing also the probability density of almost-zero angles although complete tangencies are in fact invariant. This is completely due to the PDF transformation law and strongly suggests that any statistical inference from such distributions must be generally taken with care.
\end{abstract}

\keywords{\\Covariant Lyapunov vectors, Oseledets splitting, hyperbolicity, invariance, Standard map.}

\newpage
\section{Introduction}
\noindent It is known that tangencies between invariant manifolds of unstable periodic points are directly connected to important dynamical properties, like mixing and transport \cite{z08}; indeed when invariant manifolds are almost tangent, the local dynamics can be slowed down and the invariant partitions globally can induce reductions of the transport flux \cite{w90}.
In Hamiltonian systems this phenomenon is localized around stable islands and brings the orbit to temporarily \emph{stick} there giving transient accumulations of probability density \cite{m94}; for this reason it is known as the \emph{stickiness} phenomenon \cite{z08}.\\
\\
These facts have a deep impact on the finite time behavior of many properties of the system (diffusion, correlations, time averages and recurrences) typically slowing down convergence (power laws instead of exponential decays) and, in extreme cases, making the numerical calculation of asymptotic quantities impossible \cite{ll92}; sensitivity to stickiness has been proved to be present for any observable Birkhoff average \cite{m09} and it can be checked e.g. for the finite time Lyapunov exponents (FTLE), whose distributions exhibit large deviations connected to a power laws decay of correlations (\cite{am09},\cite{omb08}).\\
\\
In spite of the known connections between full tangencies and stickiness it remains to be understood which information can be extracted from the splitting angle statistics; for this reason we focus on the possible ambiguity of such non-invariant object. The paper is organized as follows: in Sec.\ref{clv} we review 
the basic definition of CLV and transversality for bidimensional maps 
 and how these quantities transform under a generic coordinate change. Numerical
results on the Standard map are then presented and discussed in Sec.\ref{nr}, where we compare splitting angle statistics in two coordinate systems connected by a linear transformation. Sec.\ref{cnc} gives a summary of the results.

\section{The covariant Lyapunov vectors (CLV)}
\label{clv}

By definition, 2D hyperbolic dynamical systems are such that the tangent space at almost every point admits a decomposition (Oseledets splitting) in two distinct linear subspaces (\emph{stable} and \emph{unstable}), implying that these must be transversal almost everywhere \cite{o68},\cite{p77}. The covariant Lyapunov vectors (CLV) are defined at a trajectory point ${\bf x}_n$ as the unit vectors ${\bf v}_n^u$ and ${\bf v}_n^s$ respectively spanning the Oseledets subspaces (the prefix \emph{covariant} refers to their intrinsic nature \cite{gptclp07}); the angle between them is given by the euclidean inner product :
\begin{align}
\varphi_n = \cos^{-1}({\bf v}_n^u\cdot{\bf v}_n^s )\ .
\label{taf}
\end{align}

This formula yields angles $\varphi\in[0,\pi]$ while in \cite{gptclp07} splitting angles were defined in $[0,\frac{\pi}{2}]$ since the absolute value of the inner product was considered. This difference is already present in other works on the subject : in \cite{kk09,bp10,bpd10} and \cite{yr08} the absolute value is used, while in \cite{ytgcr09,tm11} it is not.
In principle hyperbolicity only requires transversality ($\varphi\neq0,\pi$), i.e. doesn't depend on whether the angle is above or below $\frac{\pi}{2}$, but, as shown in the next sections, numerical observations suggest that such distinction clarifies the comparison between different coordinate systems and seems to be the correct one.\\
\\
It is implicitly understood that collecting CLV splitting angles along chaotic trajectories interspersed through the folds of periodic points invariant manifolds yields information about the structure of those invariant manifolds themselves \cite{gptclp07}; this is because Oseledets splitting aligns in some sense to the invariant curves that cover densely the region containing the chaotic orbits \cite{p77}.\\
\\
Whichever is the case under study, applying a smooth coordinates transformation may simultaneously induce dilations and contractions of the angles, also even widening small angles while squeezing larger ones; one can realize that for a linear transformation this deformations depends on the direction of the CLV pair only, while for a nonlinear one the CLV phase space localization would also come into play.\\
To better probe this effect consider the $\mathcal{C}^1$ transformation $\tau$ from coordinates $\bf x$ to $\hat{\bf x}$ (we consider now two-dimensional systems), with Jacobian matrix $\bf M$, mapping  from the associated covariant vectors $\textnormal{d}\bf x$ to $\textnormal{d}\hat{\bf x}$ :
\begin{equation}
	\bf x\ \ \ \mapsto\ \ \ \hat{\mathbf{x}}=\tau(\mathbf{x})\ \ \ ,\ \ \ \textnormal{d}\bf x\ \ \ \mapsto\ \ \ \textnormal{d}\hat{\bf x}=\bf M\textnormal{d}\bf x
\end{equation}
One can then define the angles $\alpha^u$, $\alpha^s$ as the slopes (w.r.t. a fixed direction) of the unitary vectors ${\bf v}^u$, ${\bf v}^s$ \emph{before} the transformation, and the angles $\hat{\alpha}^u$, $\hat{\alpha}^s$ as the slopes of ${\hat{\bf v}}^u$, ${\hat{\bf v}}^s$ \emph{after} the transformation; splitting angles are then, respectively :
\begin{equation}
	\varphi\equiv(\alpha^s-\alpha^u)\ \ \ \stackrel{\tau}{\rightarrow}\ \ \ \hat{\varphi}\equiv(\hat{\alpha}^s-\hat{\alpha}^u)
\end{equation}
From the Jacobian matrix $\bf M$ it is possible to define the symmetric matrix :
\begin{equation}
	\left(
\begin{array}{cc}
	a&b\\
	b&d
\end{array}\right)\ \equiv\ {\bf M}^T\bf M
\end{equation}
and derive the useful relation :
\begin{equation}
\cot(\hat{\varphi})=\frac{1}{\det\bf M}\left[\left(\tfrac{a+d}{2}\right)\cot(\varphi) + \frac{\left(\left(\tfrac{a-d}{2}\right)\cos(\theta)+b\sin(\theta)\right)}{\sin(\varphi)}\right]\ \ \ ,\ \ \ \theta\equiv\alpha^s+\alpha^u
\label{tra}
\end{equation}
with $\theta/2$ the ``average'' direction of the CLV pair before the transformation; then under \emph{linear} changes of coordinates (for which $a,b,d$ are constants) the new splitting angle is a \emph{nonlinear} function $\hat{\varphi}=f(\varphi,\theta)$, while for nonlinear changes the coefficients $a,b,d$ are functions of the phase space (old) coordinates $\mathbf{x}$ and thus $\hat{\varphi}=f(\varphi,\theta,\mathbf{x})$. As one can expect even in the linear case the new splitting angle $\hat{\varphi}$ depends nonlinearly on the old $\varphi$ and $\theta/2$.\\
These imply that a numerical measure of transversality between invariant directions is a priori ambiguous, and any quantity depending on it may be coordinate dependent; on the other hand invariant quantities do exist, the most celebrated example being the \emph{Lyapunov exponent} : any deviation of the FTLE induced by a coordinates change asymptotically goes to zero as the inverse time \cite{ll92}. This means that non-invariance of FTLE eventually disappears in the infinite time limit, but also that any finite time observation must be regarded as coordinates representation dependent; while this can be of little implication in a single orbit observation, it is not yet clear which are its effects on ensemble observations, e.g. in analyzing FTLE distributions (as in \cite{am09}) in different coordinate systems.

\section{Numerical Results}
\label{nr}

\begin{figure}[ht!]
\centering
		\psfig{file=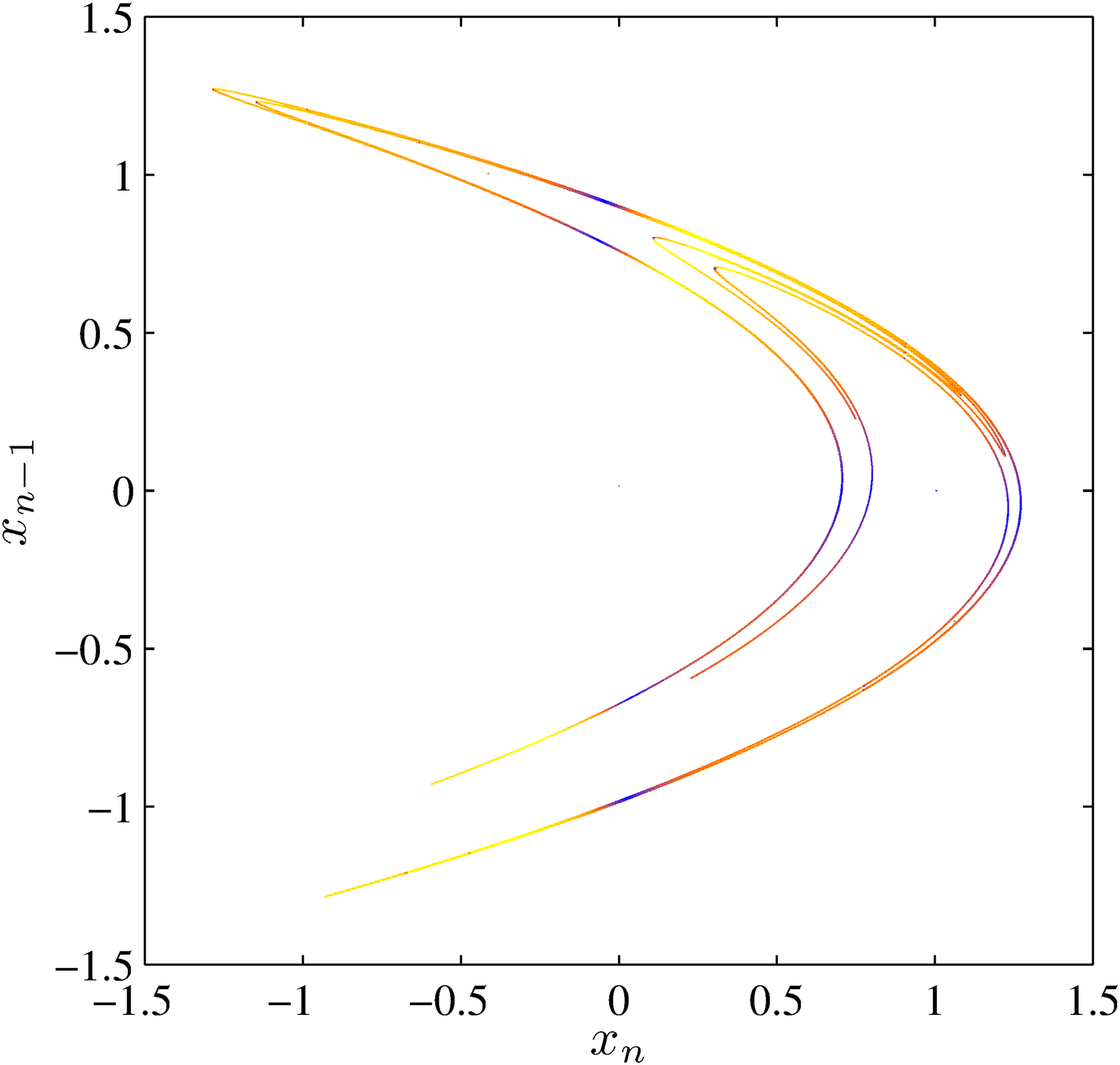,width=0.45\linewidth}
		\psfig{file=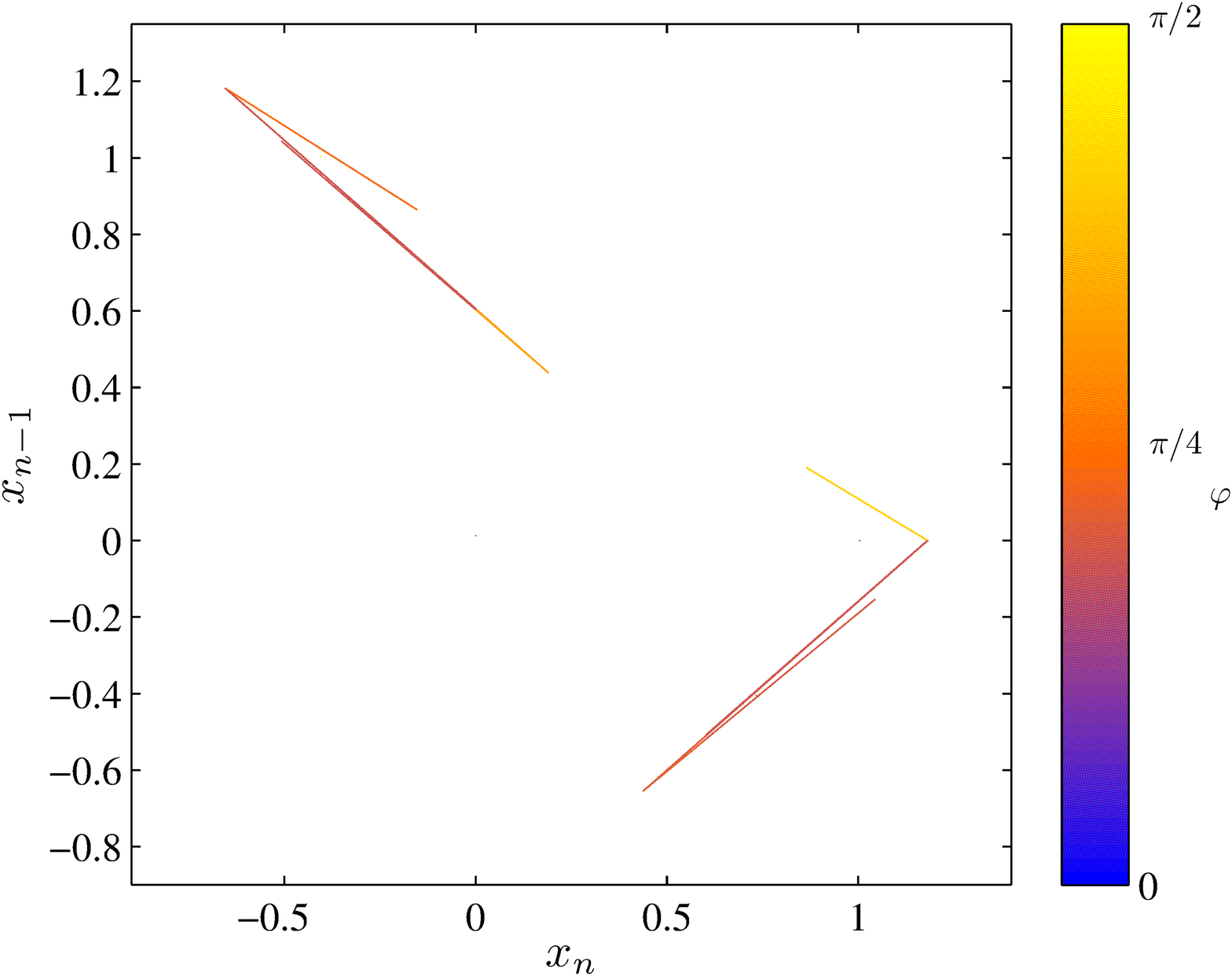,width=0.45\linewidth}
	\caption{Phase space for the H\'enon (left) and Lozi (right) map, $a=1.4\ ,\ b=0.3$. In each point of a single $10^5$ iterations orbit, transversal angle is represented by a color from palette (yellow$\ =\pi/2$, orange$\ =\pi/4$, blue$\ =0$).}
\label{helo}
\end{figure}

We first study the H\'enon and Lozi maps, \cite{h76},\cite{jps92}, both at fixed parameter values $a=1.4$ and $b=0.3$, so each one is a \emph{dissipative} system for which a strange attractor does exist. Since forward iteration of these maps brings orbits to evolve either to infinity or onto the attractor depending on initial conditions, the only points from which it is possible to extract and localize information are those belonging to the attractor. We check the correspondence with analogous calculations in \cite{gptclp07}; since the procedure described there is necessary only in more than two dimensions, the employed algorithm has been adapted and optimized for 2D maps \cite{as11}. In Fig.(\ref{helo}) the angles between stable and unstable manifolds are shown through colors from palette in each of the orbit points.\\
\\
To illustrate the action of a linear transformation we then consider the Chirikov-Taylor standard map in two appropriate coordinate systems; for each system we compute probability distributions of angles for a scan over several nonlinearity parameter values and then visualize the associated phase space distributions only for four parameter values. In these analysis we obtain angles by the formula from \cite{gptclp07} (i.e. getting angles $\varphi\in[0,\frac{\pi}{2}]$), as one in principle would do in searching for zero angles only. We then compute again the probability distributions for the four nonlinearity parameters by using Eq. (\ref{taf}), so that $\varphi\in[0,\pi]$, and we show that the angles PDF seem to be periodic in $\varphi$, with period equal to $\pi$, suggesting that splitting angles should be computed over such a range.


\subsection{Chirikov-Taylor map}
\begin{figure}[ht!]
  \centering
  \psfig{file=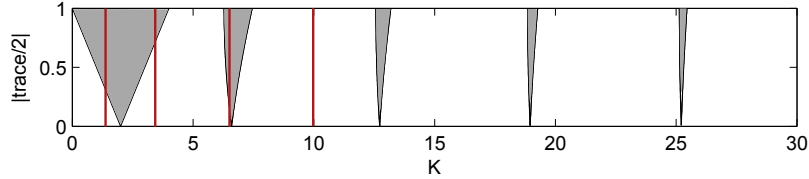,width=0.7\linewidth}
  \caption{Nonlinearity parameter $K$ windows (light gray) in which stable fixed points exist, i.e. $|\textnormal{trace}/2|<1$. The values under study range over $K\in[0,10]$; red lines mark the  values, $K=1.5,3.5,6.5,10 $, to which Fig. from (\ref{sm1.5}) to (\ref{sm10}) refer.}
\label{par}
\end{figure}

\begin{figure}[ht]
	\centering
		\psfig{file=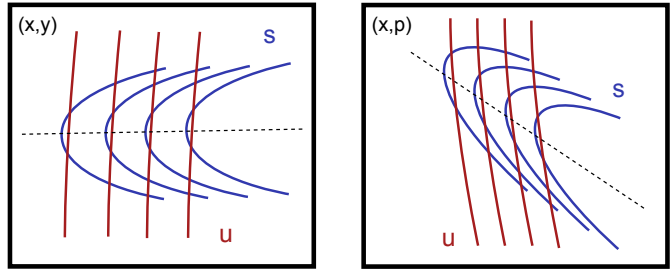,width=0.55\textwidth}
	\caption{Schematic structure of the invariant manifolds intersections in the two coordinate systems ; $u$ unstable (orange), $s$ stable (blue) manifolds. The dashed lines correspond to tangencies.}
	\label{draw}
\end{figure}

\begin{figure}[ht!]
	\centering\vspace{10pt}
		\psfig{file=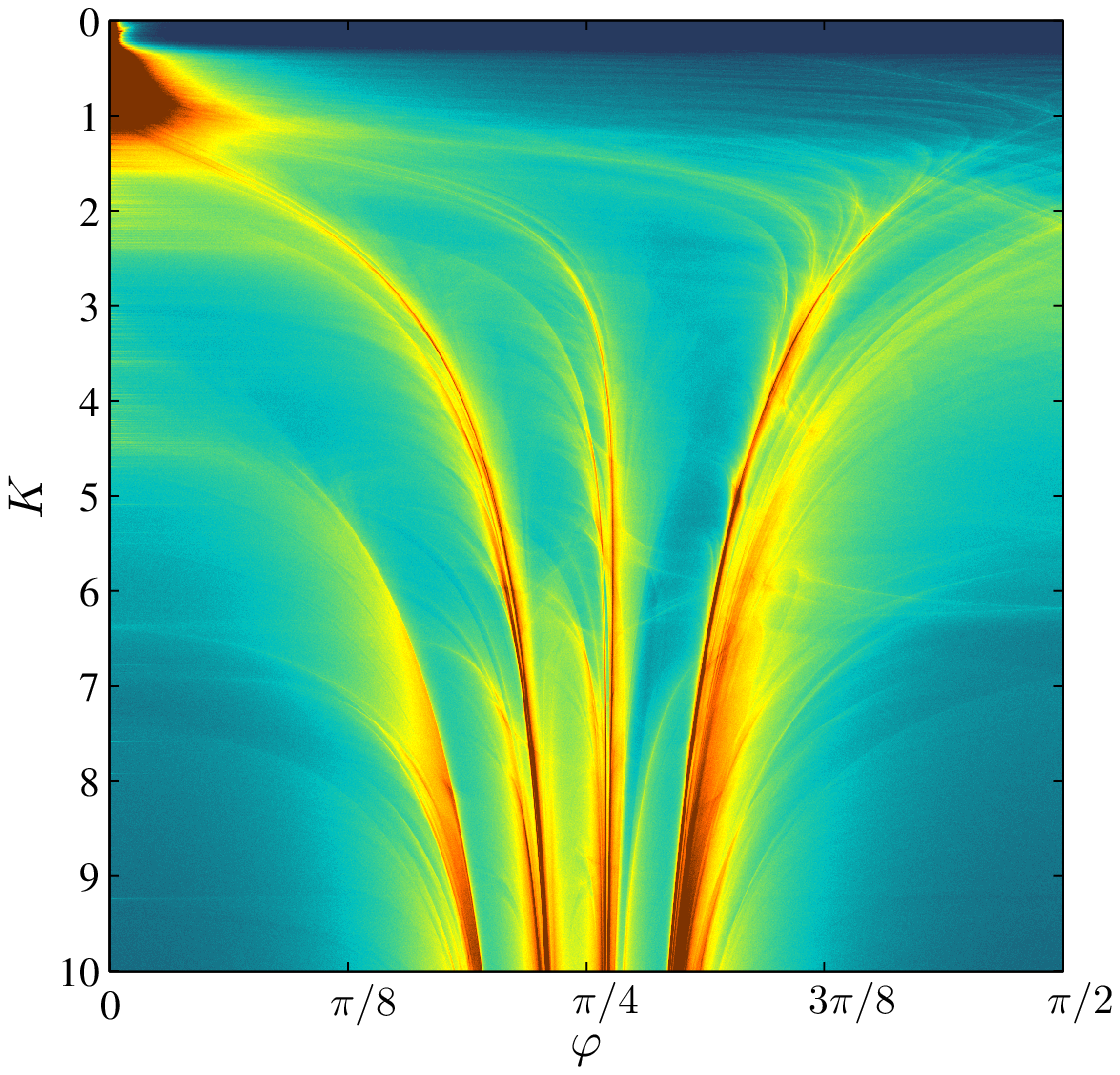,width=.615\linewidth}\vspace{10pt}
		\psfig{file=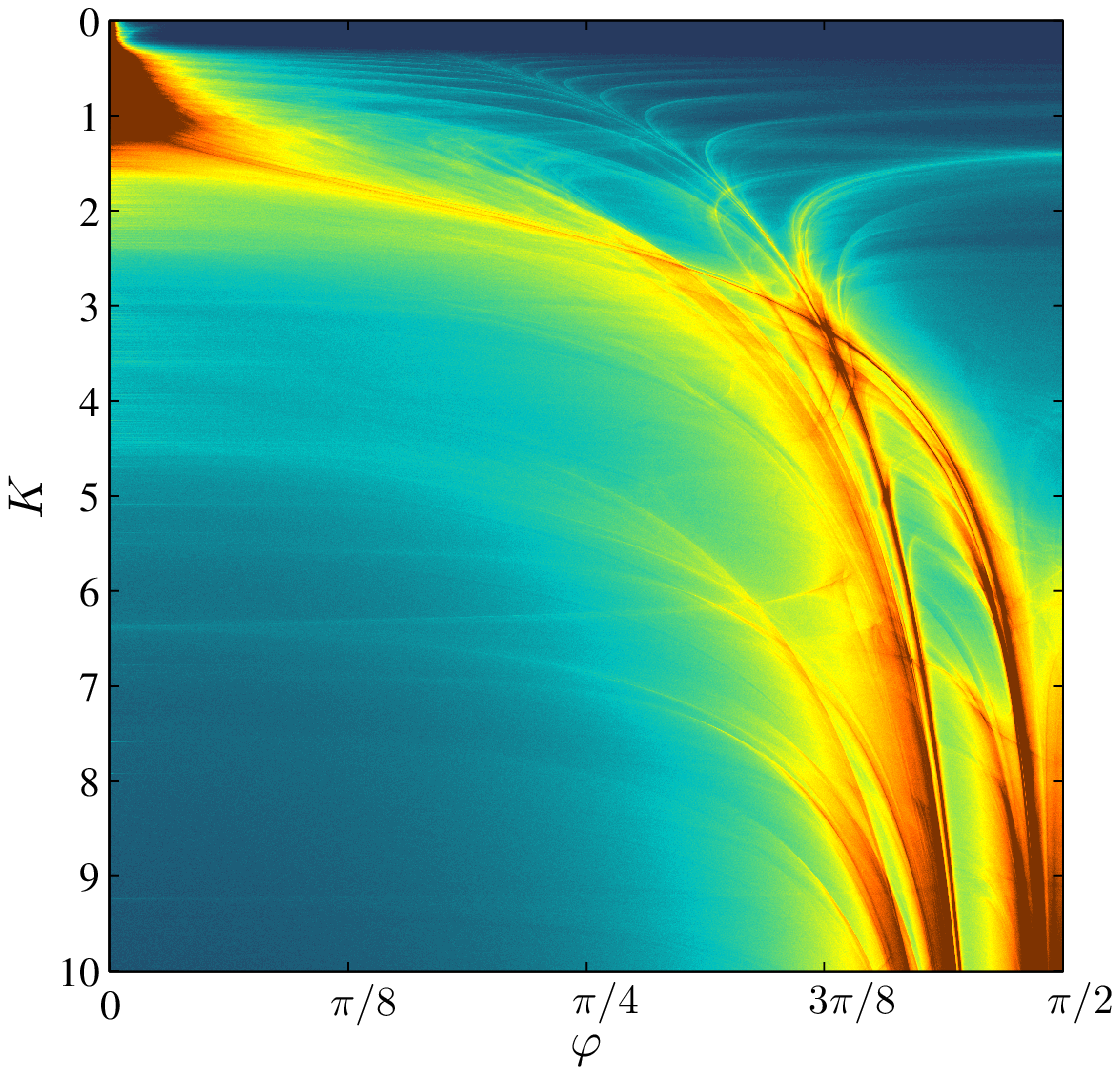,width=.615\linewidth}\vspace{10pt}
	\caption{Probability distribution function $P(K,\varphi)$ in colormap (arbitrary scale) as a function of the angles $\varphi$ and the nonlinearity parameter $K$ for the two coordinate systems ($(x,y)$ top, $(x,p)$ bottom); notice that the angles domain is $[0,\frac{\pi}{2}]$, and that some structures in the histograms seem to overlap across the upper plane.}
	\label{dis-sm}
\end{figure}

\begin{figure}[ht!]
\centering
\vspace{10pt}
\psfig{file=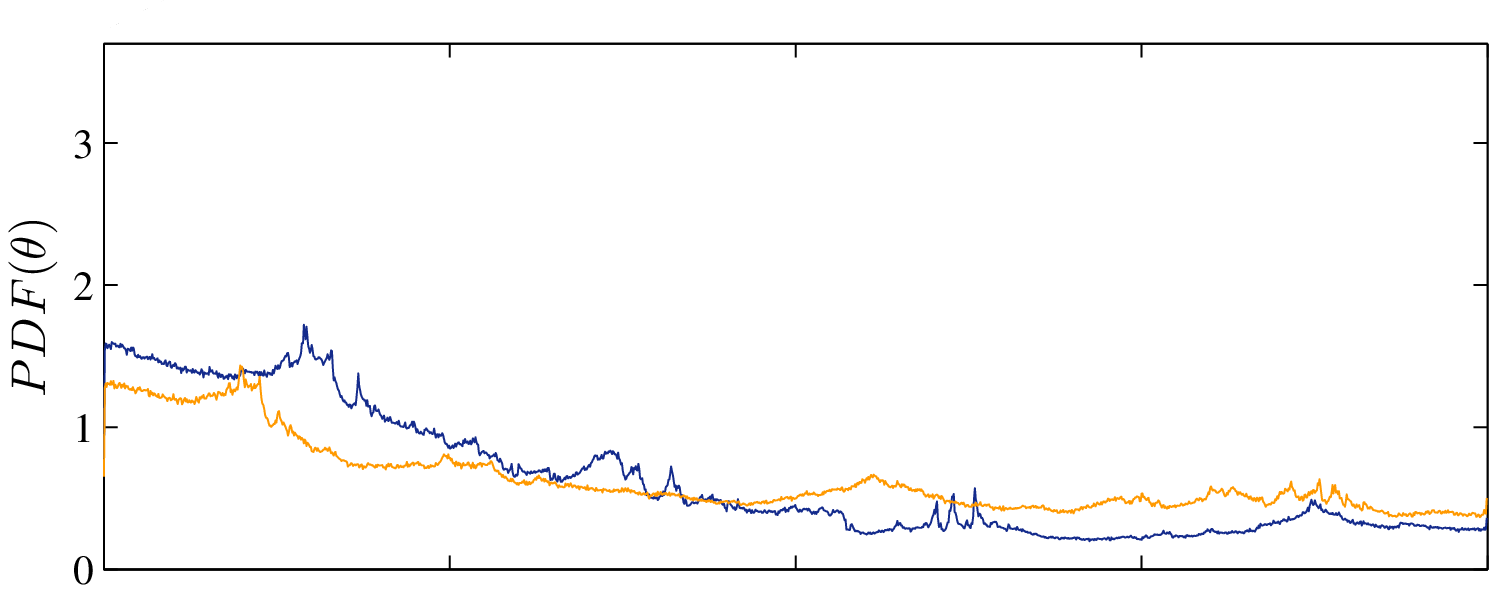,width=.8\linewidth}\vspace{-20pt}
\psfig{file=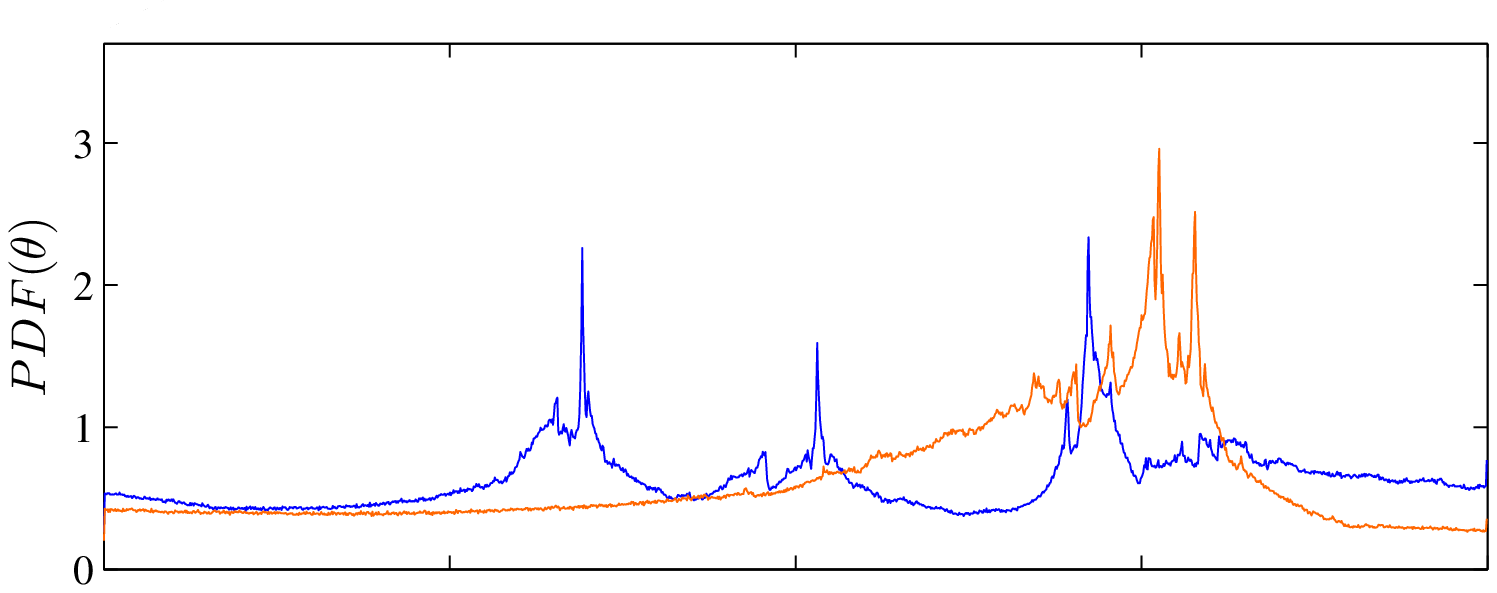,width=.8\linewidth}\vspace{-20pt}
\psfig{file=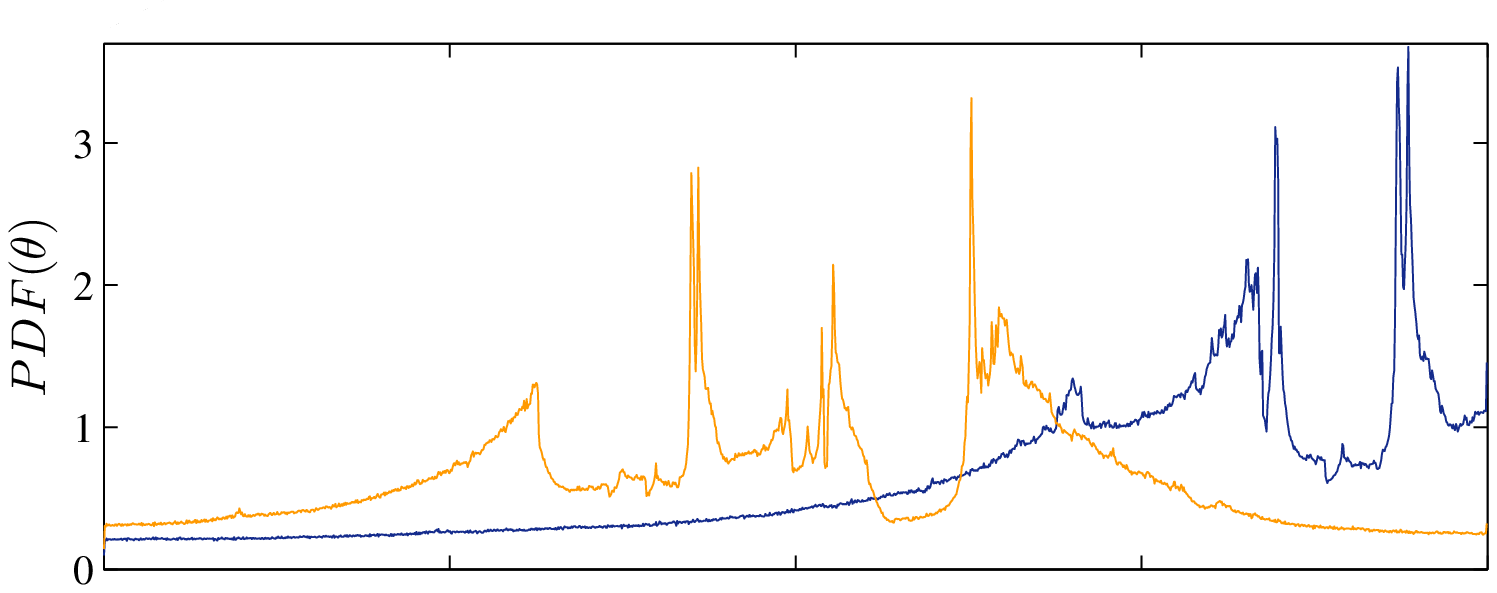,width=.8\linewidth}\vspace{-20pt}
\psfig{file=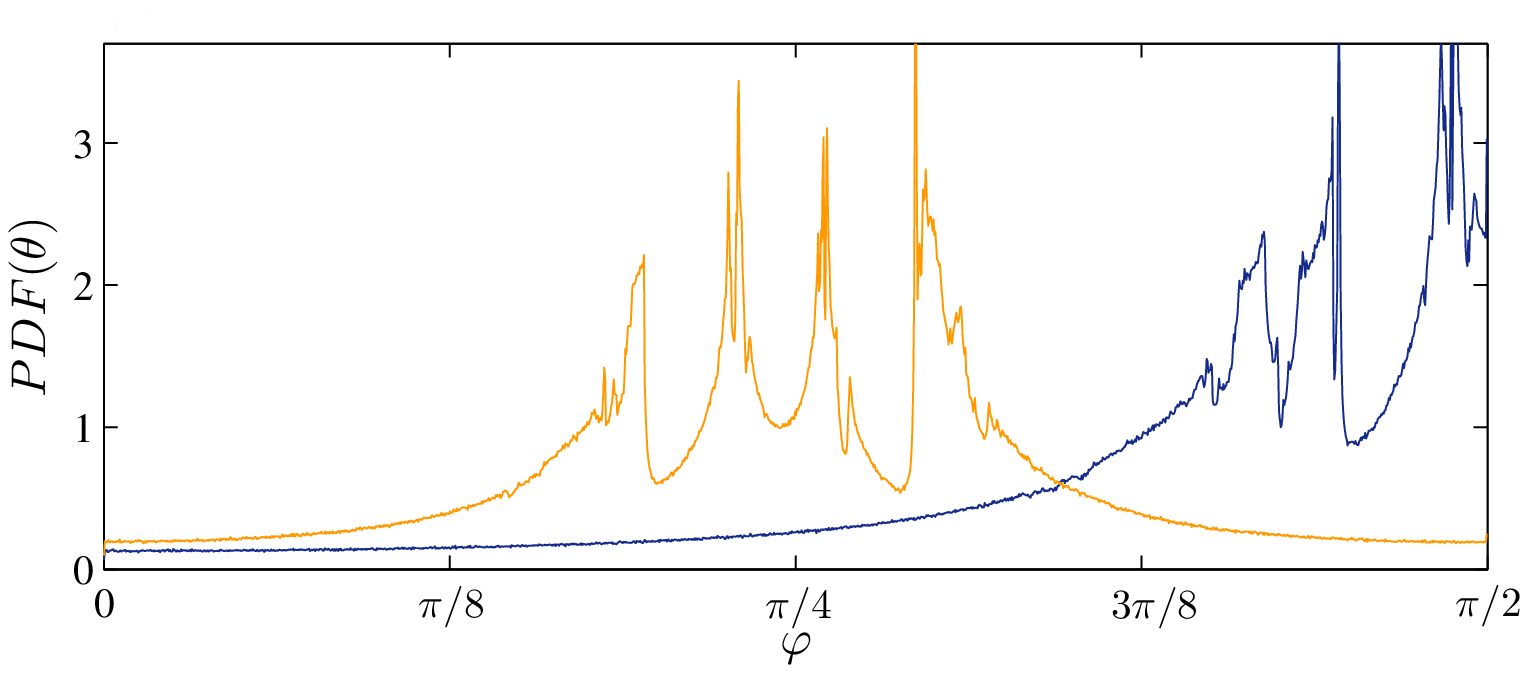,width=.8\linewidth}\vspace{10pt}
\caption{Probability distribution function of splitting angles absolute values for the Standard map 
in the $(x,y)$ (blue/black) and $(x,p)$ (orange/gray) coordinates for parameter values 
$K = 1.5\ (a),\ 3.5\ (b),\ 6.5\ (c),\ 10\ (d)$ (vertical axis in arbitrary scale).}
\label{his}
\end{figure}

\begin{figure}[ht!]
\centering
\psfig{file=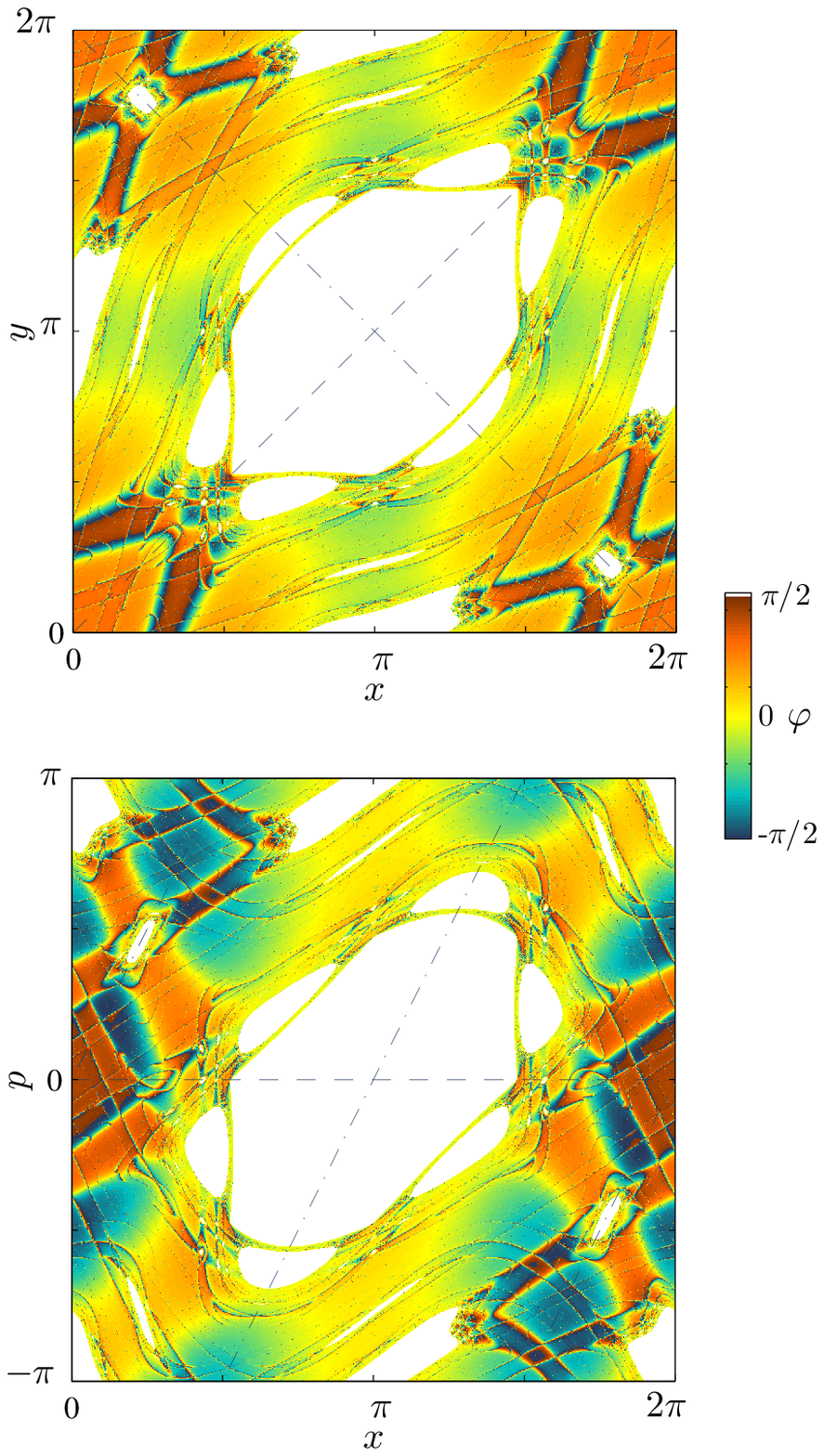,width=.7\linewidth}\vspace{10pt}
\caption{Phase space for the standard map, $K=1.5$, in the $(x,y)$ coordinates (top), 
in the $(x,p)$ coordinates (bottom). In each point of a single $10^9$ 
iterations orbit, splitting angle is represented by a color from palette (orange=$\pi/2$, yellow$=0$, blue$=-\pi/2$); dashed and dot-dashed lines are the square diagonals in the $(x,y)$ plane and their images in the $(x,p)$ plane.}
\label{sm1.5}
\end{figure}

\begin{figure}[ht!]
\centering
\psfig{file=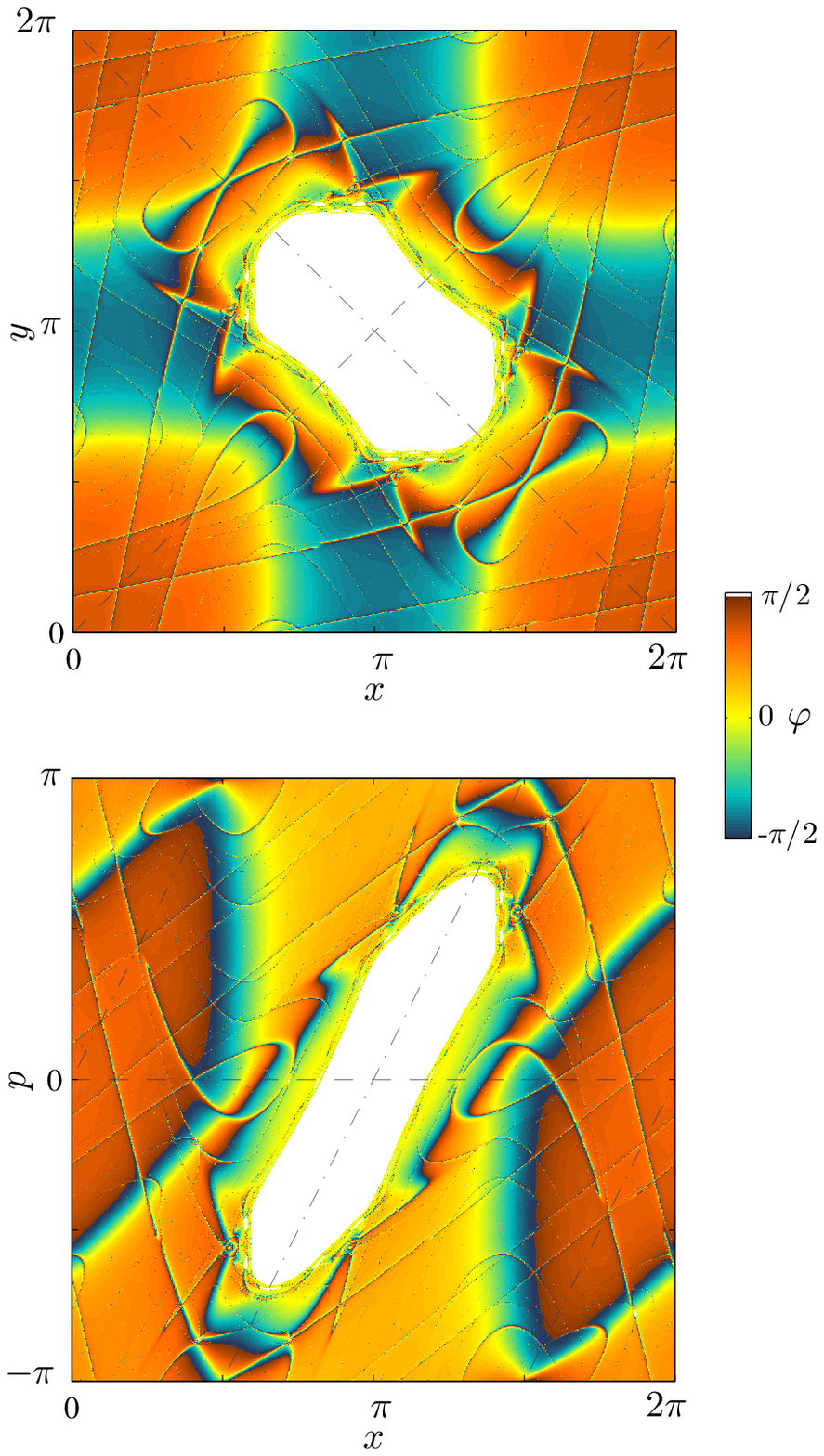,width=.7\linewidth}\vspace{10pt}
\caption{Phase space for the standard map, $K=3.5$, in the $(x,y)$ coordinates (top), 
in the $(x,p)$ coordinates (bottom). In each point of a single $10^9$ 
iterations orbit, splitting angle is represented by a color from palette (orange=$\pi/2$, yellow$=0$, blue$=-\pi/2$); dashed and dot-dashed lines are the square diagonals in the $(x,y)$ plane and their images in the $(x,p)$ plane.}\label{sm3.5}
\end{figure}

\begin{figure}[ht!]
\centering
\psfig{file=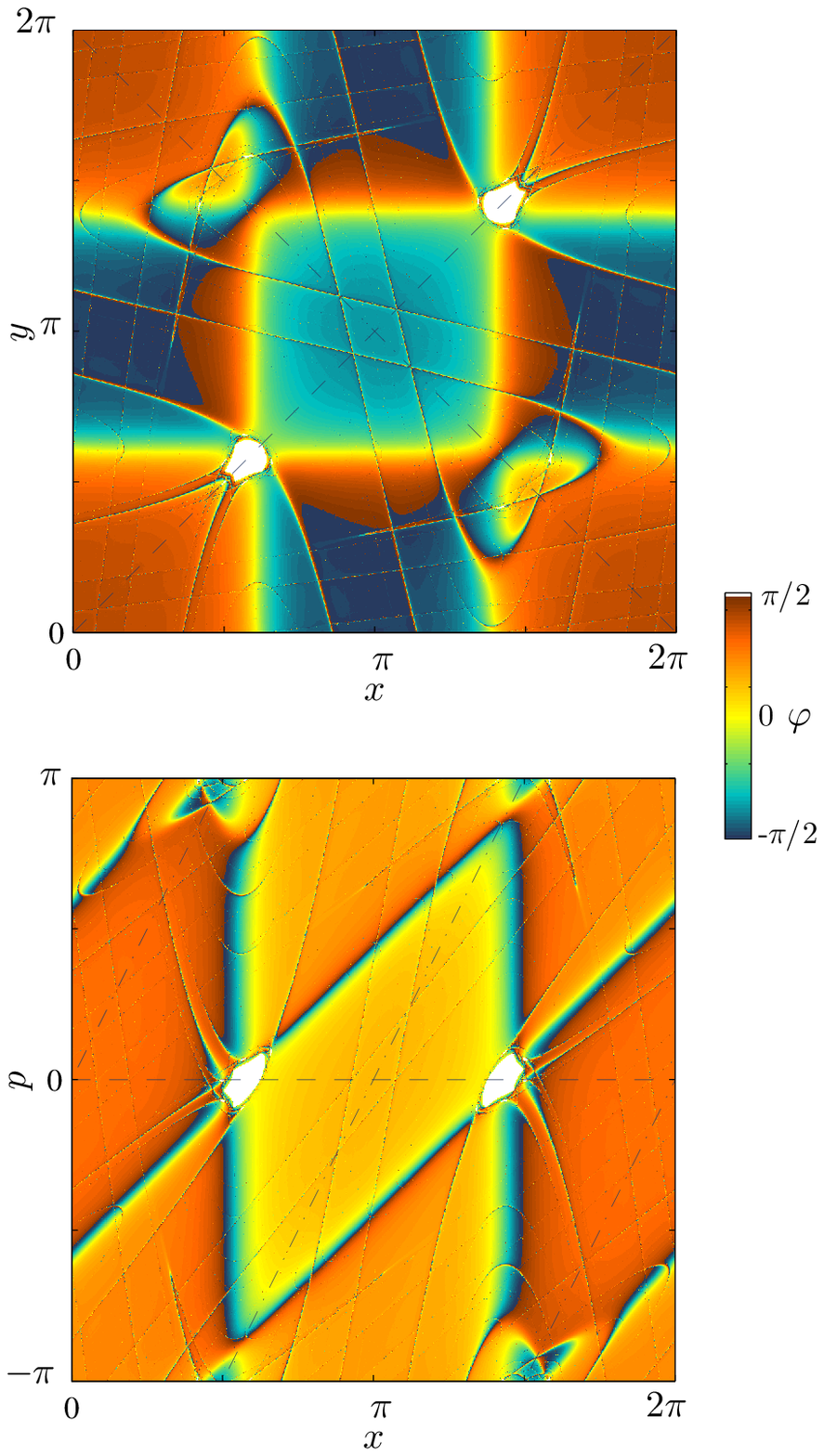,width=.7\linewidth}\vspace{10pt}
\caption{Phase space for the standard map, $K=6.5$, in the $(x,y)$ coordinates (top), 
in the $(x,p)$ coordinates (bottom). In each point of a single $10^9$ 
iterations orbit, splitting angle is represented by a color from palette (orange=$\pi/2$, yellow$=0$, blue$=-\pi/2$); dashed and dot-dashed lines are the square diagonals in the $(x,y)$ plane and their images in the $(x,p)$ plane.}\label{sm6.5}
\end{figure}

\begin{figure}[ht!]
\centering
\psfig{file=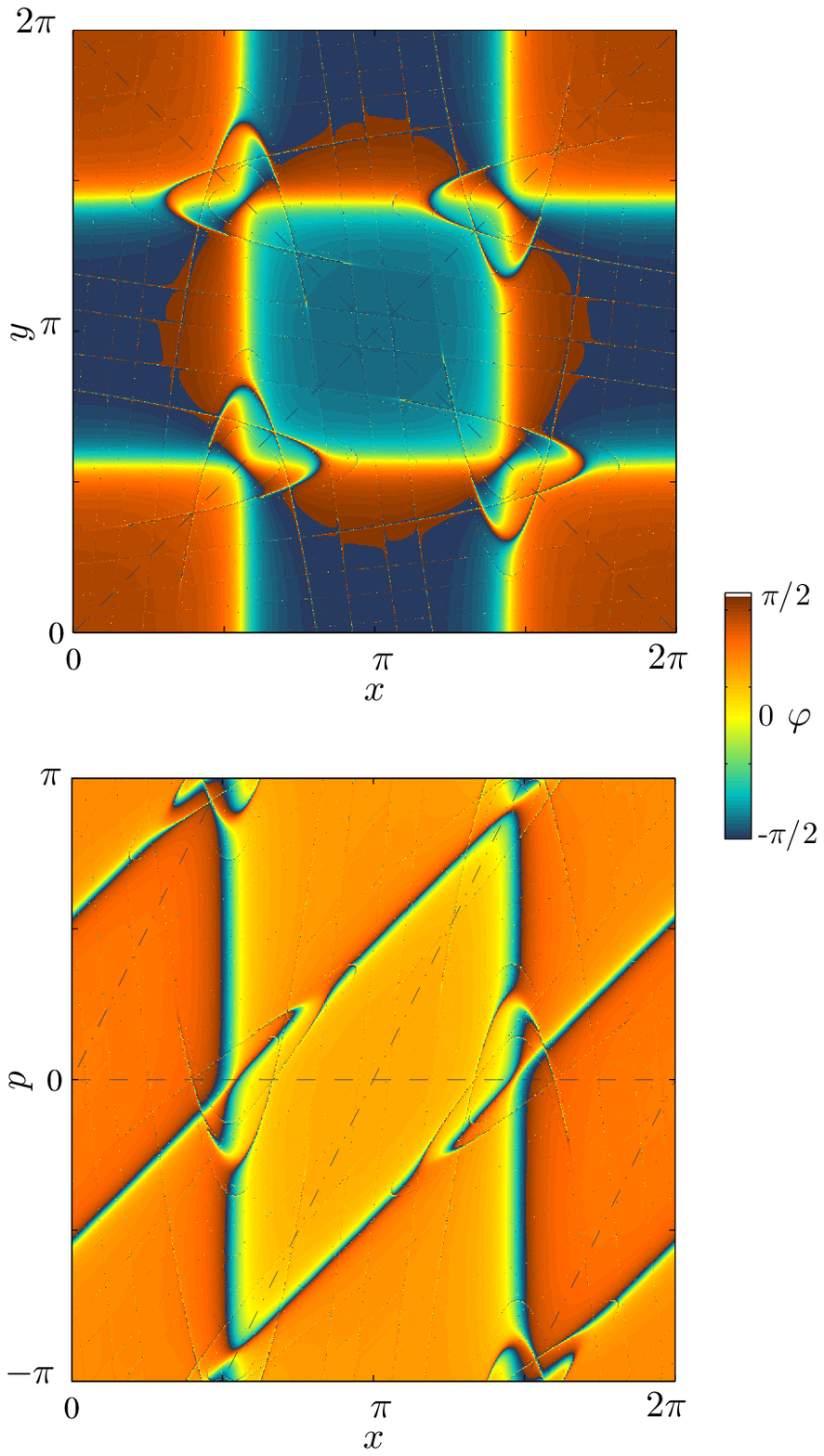,width=.7\linewidth}\vspace{10pt}
\caption{Phase space for the standard map, $K=10$, in the $(x,y)$ coordinates (top), 
in the $(x,p)$ coordinates (bottom). In each point of a single $10^9$ 
iterations orbit, splitting angle is represented by a color from palette (orange=$\pi/2$, yellow$=0$, blue$=-\pi/2$); dashed and dot-dashed lines are the square diagonals in the $(x,y)$ plane and their images in the $(x,p)$ plane.}\label{sm10}
\end{figure}

The Chirikov-Taylor standard map, being the paradigm of all chaotic Hamiltonian maps, is a good benchmark to test hyperbolicity properties. In canonical coordinates it has the form :
\begin{align}
&\nonumber x_{n+1}\ =\ x_n\ +\ p_{n+1}\quad \mathrm{mod} \, 2\pi\\
&p_{n+1}\ =\ p_n\ +\ K\sin(x_n)
\label{sm}
\end{align}
with $K$ the nonlinearity parameter; it is possible to consider also the another representation $(x_n,x_{n-1})$, by defining $y\equiv (x-p)$ i.e. $y_n\equiv x_{n-1}$: this leads to the following form:
\begin{align}
&\nonumber x_{n+1}\ =\ 2x_n+\ K\sin(x_n)-y_{n}\quad \mathrm{mod} \, 2\pi\\
&y_{n+1}\ =\ x_n
\label{sm2}
\end{align}
This evolution has the nice property of being reversed by coordinates exchange, i.e. reflection w.r.t. the bisector is equivalent to time inversion; on the other hand it has the drawback of loosing information about momentum growth since both coordinates belong to the interval $[0,2\pi]$. The transformation that brings from coordinates $(x,p)$ to $(x,y)$ is linear and thus coincides with its Jacobian matrix :
\begin{equation}
	\tau\sim\bf M\ =\ \left(
\begin{array}{rr}
1&0\\
1&-1
\end{array}
\right)\ \ \ \Rightarrow\ \ \ {\bf{M}}^T\bf{M}\ =
\ \left(
\begin{array}{rr}
2&-1\\
-1&1
\end{array}
\right)\ =\ 
\left(
\begin{array}{cc}
a&b\\
b&d
\end{array}
\right)\ \ ,\ \ \det\bf M\ =\ -1\ .
\label{tra1}
\end{equation}
We remark that this transformation is an involution (i.e. coincides with its inverse); the transformation law (\ref{tra}) for transversal angles associated to (\ref{tra1}) then reads :
\begin{equation}
\cot(\hat{\varphi})=-\frac{3}{2}\cot(\varphi) + \frac{\sin(\theta)-\tfrac{1}{2}\cos(\theta)}{\sin(\varphi)}
\label{traSM}
\end{equation}
remember that $\theta\equiv\alpha^s+\alpha^u$ is the CLV pair orientation; nonlinearity is apparent here, together with the $\theta$ dependence. The Chirikov-Taylor standard map possesses a very complicated phase space structure in both coordinate systems (\ref{sm}) and (\ref{sm2}), with coexistence of regular and chaotic motion for a large set of nonlinearity parameter values; we first consider a scan of $10^3$ equidistant values for the nonlinearity parameter inside the interval $K\in[0,10]$; for each coordinate system and each $K$ value we compute an ensemble of $10^3$ orbits of length $10^4$ and compute histograms of transversal angles over the set $\varphi\in[0,\frac{\pi}{2}]$ (i.e. as in \cite{gptclp07}). We also verify that computing transversality by direct calculation (over the two systems) and through the transformation law (from one to the other) is completely equivalent. The chosen $K$ interval contains elliptic fixed points and thus the associated bifurcations also; in Fig. (\ref{par}) we show the absolute value of the half trace of the Jacobian matrix (of (\ref{sm}) or (\ref{sm2}), it is an invariant) versus nonlinearity parameter $K$. It determines the Jacobian matrix eigenvalues and so, when it is less than $1$, it gives the frequency $\omega$ of elliptic fixed points via the relation :
\begin{equation}
\omega\ =\ \arccos\left|\frac{\textnormal{trace}}{2}\right|.
\end{equation}
\\
Outside windows with $|\textnormal{trace}/2|<1$ islands also exist and are associated to at least period-2 fixed points, which have a more complex localization in parameter space, usually connected to bifurcation sequences. Fig. (\ref{dis-sm}) show the PDF $P(K,\varphi)$ in colormap (arbitrary scale) as a function of the angles $\varphi$ and the nonlinearity parameter $K$ for the two coordinate systems ($(x,y)$ top, $(x,p)$ bottom). \\There
it can be noticed that some patterns across the $(\varphi,K)$ plane
seem to overlap as in a superposition of two trends w.r.t. $K$: this might be
a suggestion about the correct domain for the variable
$\varphi$. Densities for $K\gtrsim 3$ are higher around 
$\varphi=\frac{\pi}{2}$ in the $(x,y)$ coordinates and around $\varphi=\frac{\pi}{4}$ in the $(x,p)$; an increase of the nonlinearity parameter pushes probability away from zero and leaves the distributions
more compact around maxima. This can be clearly seen in Fig. (\ref{his})
where a choice of four $K$ values is extracted from 
Fig. (\ref{dis-sm}) ($K=1.5,3.5,6.5,10$) and superimposed for the
two coordinate systems; due to nonlinearity in Eq.(\ref{traSM}) the
two corresponding PDF in all of the four cases are different around $\varphi=0$, with a discrepancy far larger than the histograms error bars.\\
\\
For the same four values of the nonlinearity parameter $K$ we
observe the splitting patterns through phase space (Figs. from
(\ref{sm1.5}) to (\ref{sm10})) by plotting in each point of an orbit a color
associated to the angle value by some color bar. Since chaotic
trajectories almost fill out areas, these plots show spatial variation
of hyperbolicity  w.r.t. nonlinearity strength: in
both representations it is apparent that stronger nonlinearities yield wider
regions of smoothly varying angles, thus leading to higher and more
compact peaks in the associated PDF from Fig. (\ref{dis-sm}) and
(\ref{his}). We remark that all these data have been collected by the formula in \cite{gptclp07} so that angles still belong to $[0,\frac{\pi}{2}]$; indeed, the main histograms features are already apparent but no clear connection between the two coordinate system is visible yet.\\
\\
In Fig. (\ref{sm1.5}) to (\ref{sm10}) it is interesting to notice how the angles go near zero (blue/black color) around the islands, as one can expect, but also along certain definite structures surely not due to islands (e.g. the blue/black stripes in Fig. (\ref{sm10})); these tangencies are due to the structure of the manifolds across those regions. Indeed one can locally describe one foliation (e.g. the stable one) as rectilinear (parallel lines) and the other one as parabolic (``parallel'' curves); the intersection between such two families of curves naturally gives rise to a line of tangency between the foliations (see the dashed line in the left panel of Fig. (\ref{draw})), that is what we observe in phase space plots. Those lines of tangencies are the cause for the angles PDF in Fig. (\ref{his}) to extend quite smoothly to their values in $\varphi=0$ also in the $K=10$ case, in which macroscopic islands seem to be absent and thus one may not expect to observe any tangency.\\

\begin{figure}[ht!]
\centering
\vspace{10pt}
\psfig{file=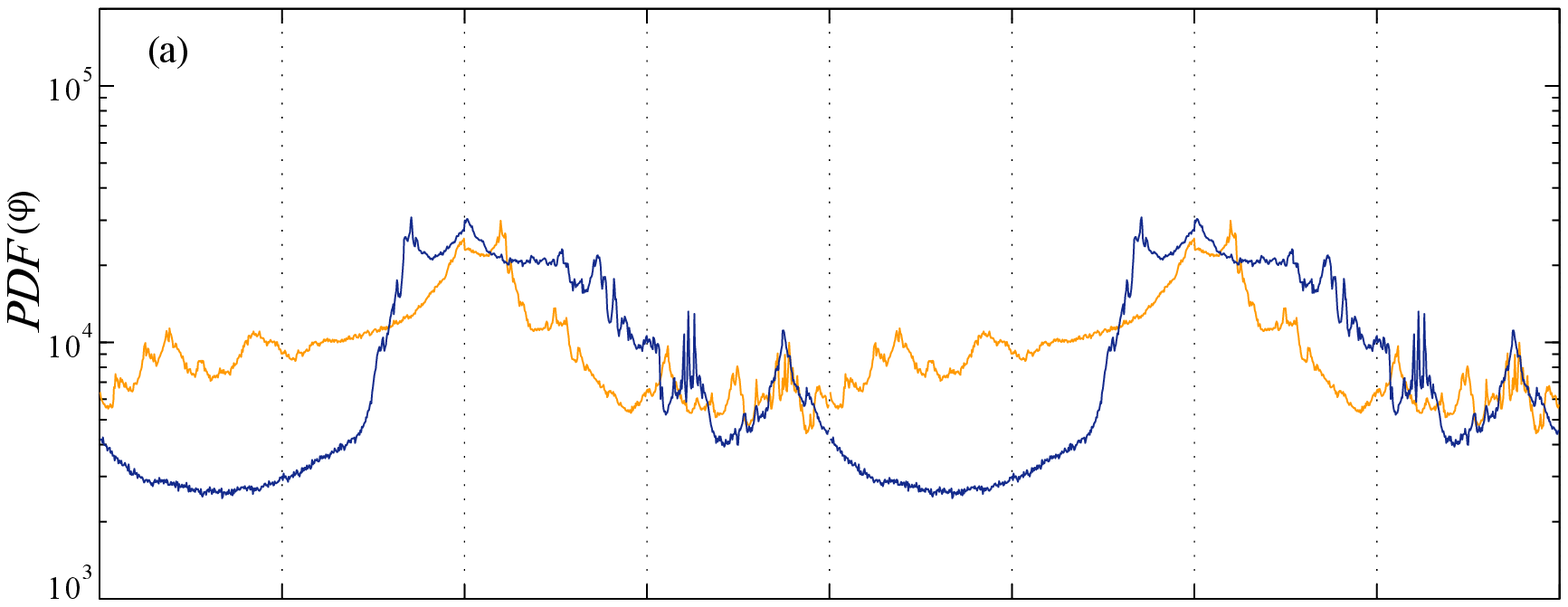,width=.775\linewidth}\vspace{10pt}
\psfig{file=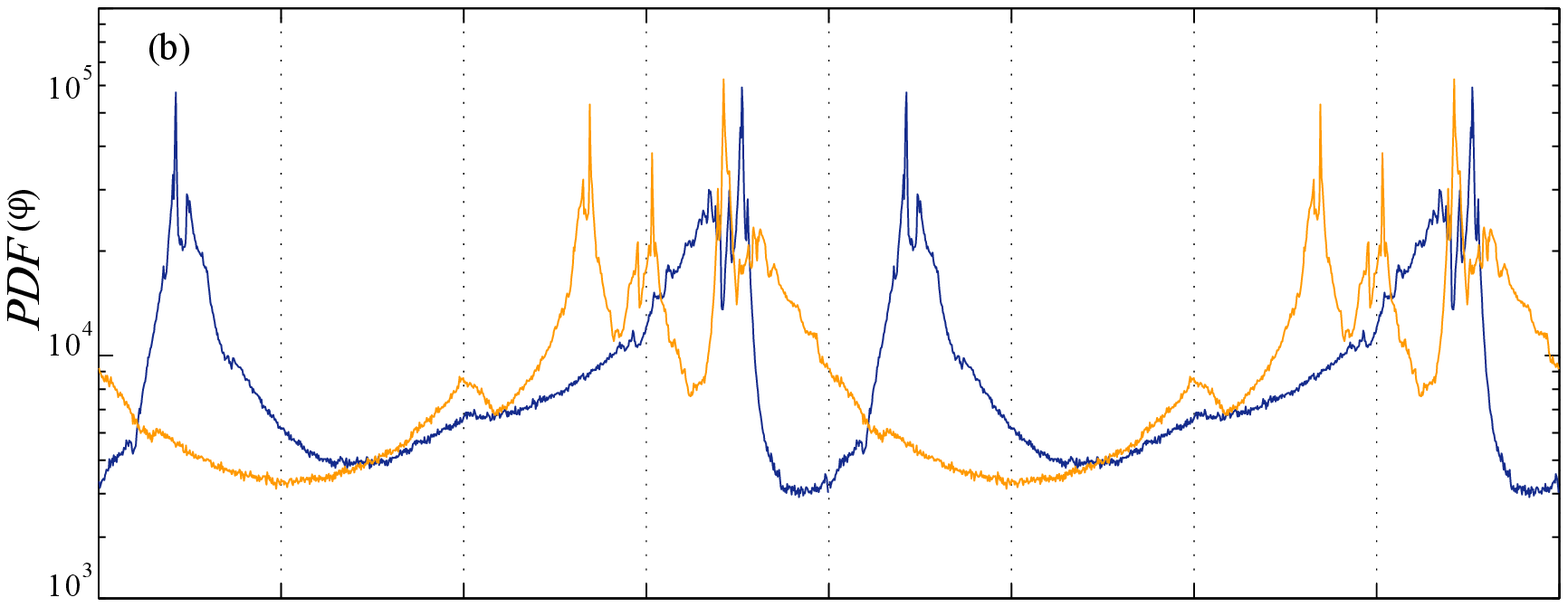,width=.775\linewidth}\vspace{10pt}
\psfig{file=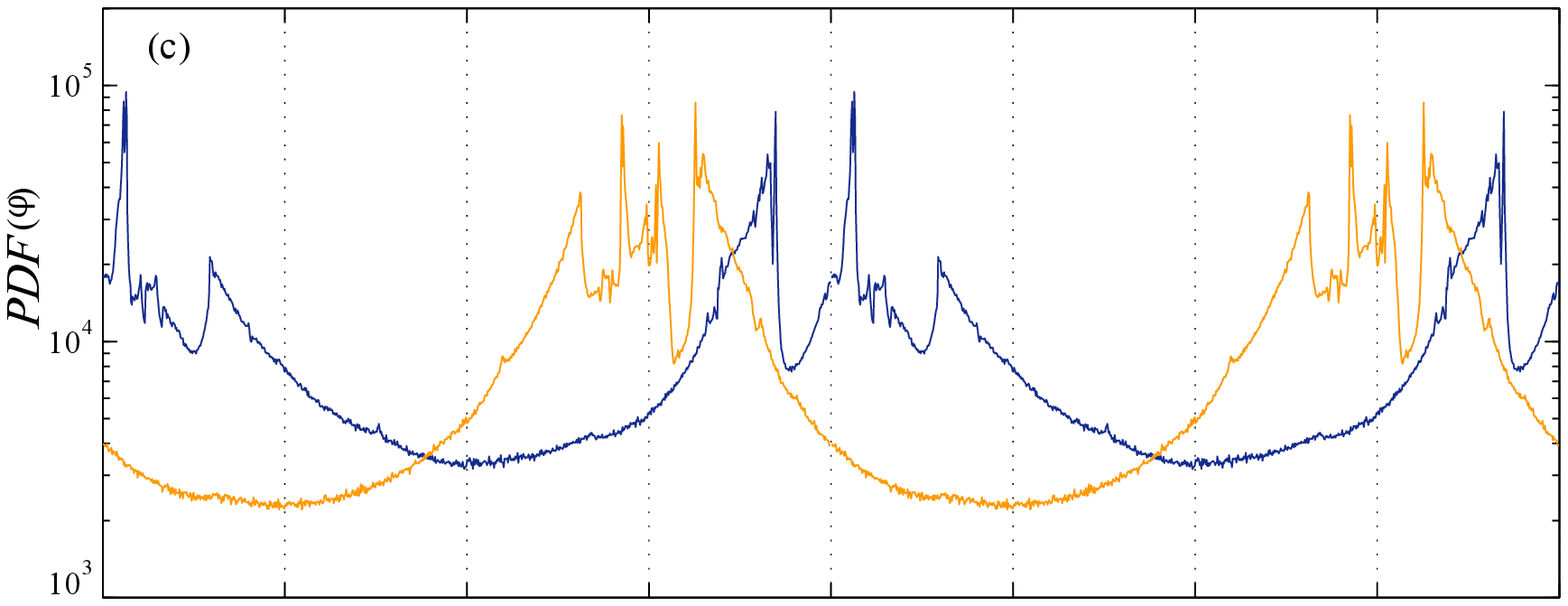,width=.775\linewidth}\vspace{10pt}
\psfig{file=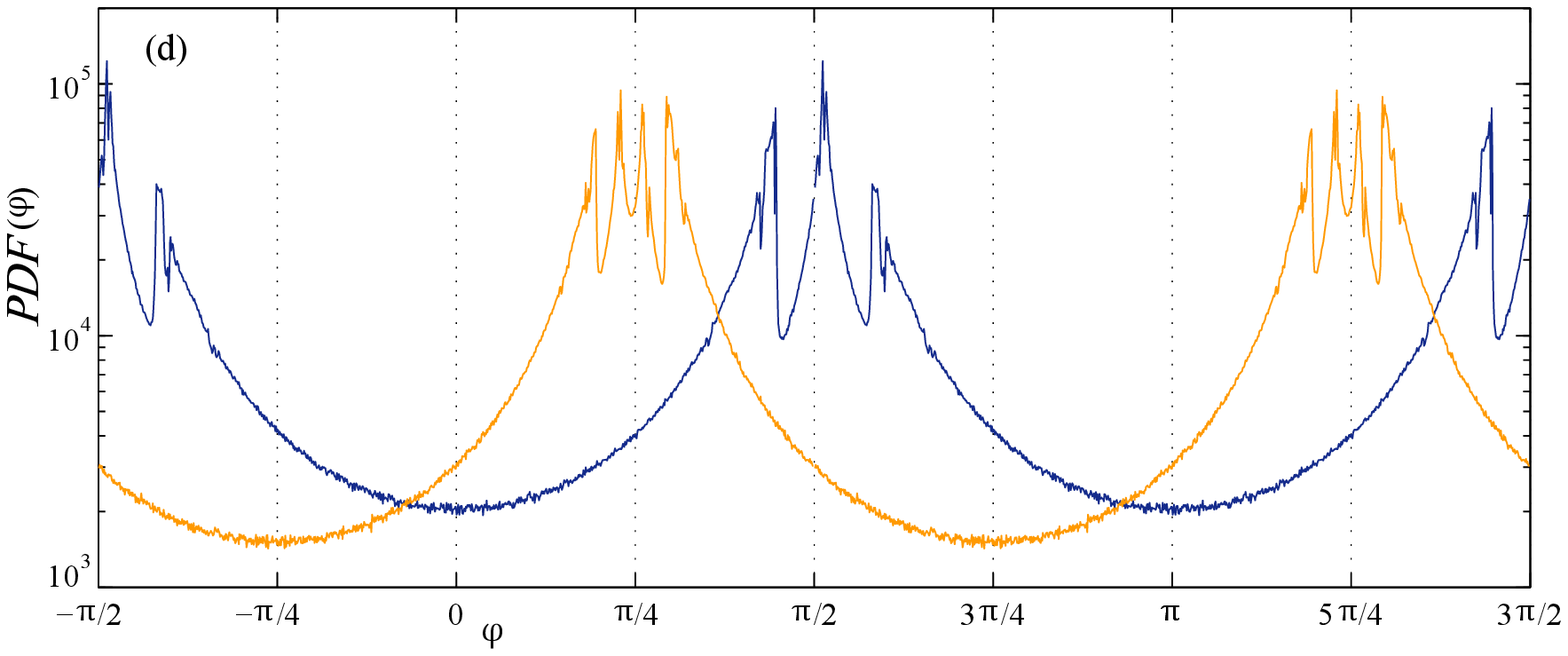,width=.775 \linewidth}\vspace{10pt}
\caption{$\log_{10}$PDF of splitting angles with sign for the Standard map 
in the $(x,y)$ (blue/black) and $(x,p)$ (orange/gray) coordinates for parameter values 
$K = 1.5\ (a),\ 3.5\ (b),\ 6.5\ (c),\ 10\ (d)$. Notice that here the distribution's domain is $[0,\pi]$ and, only for these plots, the horiziontal axis extends over $[-\frac{\pi}{2},\frac{3\pi}{2}]$ to show the natural periodicity.}
\label{hisign}
\end{figure}

What may worry about the PDF in Fig. (\ref{his}) is that, for each value of $K$, details between coordinate systems are different. Getting the angles inside the interval $[0,\pi]$ yields the histograms in Fig. (\ref{hisign}), that correspond to Fig. (\ref{his}) but extends from $-\frac{\pi}{2}$ to $\frac{3\pi}{2}$ and has log$_{10}$ vertical axis; it can be noticed that PDF are periodic w.r.t. $\varphi$ with period equal to $\pi$, so the distributions in Fig. (\ref{hisign}) are repeated twice. Qualitative connections between the two systems are now apparent: the histogram in one system is similar to the other one after a translation with inversion of the variable, as Eq. (\ref{traSM}) suggests. Moreover, the different value of the PDF in $\varphi=0$ for the two systems is completely explained by the transformation law of probability density functions : it is straightforward to obtain the relation :
\begin{align}
	\left.\frac{\partial P}{\partial \varphi}\right|_{\varphi=0}\ =\ \left.\frac{\partial\hat{P}}{\partial\hat{\varphi}}\right|_{\hat{\varphi}=0}\left.\frac{\partial\hat{\varphi}}{\partial \varphi}\right|_{\varphi=0},
\end{align}
from the invariance condition $\textnormal{d}\hat{P}(0)=\textnormal{d}P(0)$; we remark two points here : \emph{exact} tangencies cannot be numerically measured since no map can transform them into nonzero angles; it must be also taken into account that the transversality deformations from a system to another depends crucially on the orientation angle $\theta$ also, so to evaluate the exact difference between the histograms the partial derivative of Eq. (\ref{tra}) or (\ref{traSM}) (evaluated in $\varphi=\hat{\varphi}=0$) should also be integrated over $\theta$ weighted by its associated density. The net effect of these properties is that at high nonlinearity parameter $K$ both the PDF values and slopes in $\varphi=0$ are different for the two systems: for $(x,y)$ there is a minimum and the slope is zero, for $(x,p)$ the slope is clearly positive (see Fig. (\ref{hisign})). This feature can also be addressed through the manifold structures visible in Fig. (\ref{sm6.5}) and (\ref{sm10}) (resp. at $K=6.5$ and $K=10$) and schematized in Fig. (\ref{draw}): the dashed lines correspond to tangencies and divide the regions in two; the parabolic foliation in $(x,y)$ (left panel) is qualitatively symmetric w.r.t. the dashed line and the corresponding PDF in Fig. (\ref{hisign}) is almost symmetric at zero. Since the $(x,p)$ coordinates (right panel) come from the $(x,y)$ ones through a shear plus inversion, parabolic foliations may not be symmetric anymore, suggesting an heuristic explanation for the PDF asymmetry around $\varphi=0$ (Fig. (\ref{hisign})).

\section{Conclusions}
\label{cnc}
In this numerical work we investigate the action of smooth coordinate change on the Oseledets splitting for dissipative (H\'enon and Lozi maps) and conservative (standard map) bidimensional systems, to understand its qualitative consequences on invariant manifolds splitting angles. After showing how these angles transform (\ref{tra}) under the choice of new coordinates we compare their statistics and remark that, to understand the structure of their probability distribution functions, angles must be collected over the range $[0,\pi]$, i.e. taking into account also the sign of the CLV inner product. Through plots of transversality across phase space and simple arguments, we show how the linear shear connecting two coordinate systems induces local changes in the angles PDF, suggesting how to interpret the global picture at least in case of strong nonlinearity. It remains to be understood the structural difference of such probability distributions between mixed and almost uniform phase spaces (e.g. between the cases at $K=6.5$ and $K=10$ resp. in Fig. (\ref{sm6.5}) and (\ref{sm10})) and further studies are required.\\
The whole numerical work is based on the algorithm in \cite{gptclp07}, but optimized for generic bidimensional maps and also extended to manifold curvatures \cite{as11}.

\section*{Aknowledgements}
This work has been partially supported by MIUR-PRIN project
"Nonlinearity and disorder in classical and quantum transport
processes". C.M. gratefully acknowledges Universit\'a degli Studi
dell'Insubria for hospitality and financial support. 

\bibliographystyle{ws-ijbc}
\bibliography{references}

\end{document}